\documentclass[intlimits,twoside,a4paper]{article}

\usepackage{amsmath,amssymb}
\usepackage{graphicx}
\usepackage{wrapfig}

\usepackage[T2A]{fontenc}
\usepackage[cp1251]{inputenc}

\usepackage[eqsecnum]{cmpj2}

\issue{2012}{15}{2}{23606}

\doinumber{10.5488/CMP.15.23606}

\title[Consistency between diffraction data and partial radial distribution functions]%
{An independent, general method for checking consistency between diffraction data and partial radial distribution functions derived from them: \\the example of liquid water}

\author[Z.~Steinczinger, L.~Pusztai]{Z.~Steinczinger\refaddr{label1}, L.~Pusztai\refaddr{label2}\thanks{E-mail: pusztai.laszlo@wigner.mta.hu}}

\authorcopyright{Z.~Steinczinger, L.~Pusztai, 2012}

\addresses{
\addr{label1} Budai Nagy Antal Secondary School, H--1121, Budapest, Anna utca 13--15., Hungary
\addr{label2} Institute for Solid State Physics and Optics, Wigner Research Centre for Physics, Hungarian Academy of Sciences, H--1525 Budapest, P.O. Box 49., Hungary}
\date{Received January 31, 2012, in final form May 7, 2012}

\begin{document}

\maketitle

\begin{abstract}
There are various routes for deriving partial radial distribution functions of disordered systems from experimental diffraction (and/or EXAFS) data. Due to limitations and errors of experimental data, as well as to imperfections of the evaluation procedures, it is of primary importance to confirm that the end result (partial radial distribution functions) and the primary information (diffraction data) are consistent with each other. We introduce a simple approach, based on Reverse Monte Carlo modelling, that is capable of assessing this dilemma. As a demonstration, we use the most frequently cited set of  ``experimental'' partial radial distribution functions on liquid water and investigate whether the 3 partials (O--O, O--H and H--H) are consistent with the total structure factor of pure liquid D$_{2}$O from neutron diffraction and  that of H$_{2}$O from X-ray diffraction. We find that while neutron diffraction on heavy water is in full agreement with all the 3 partials, the addition of X-ray diffraction data clearly shows problems with the O--O partial radial distribution function. We suggest that the approach introduced here may also be used to establish whether partial radial distribution functions obtained from statistical theories of the liquid state are consistent with the measured structure factors.
\pacs 61.20.-p, 61.25.-f, 61.05.fm
\end{abstract}

\section{Introduction}

Experimentally determining the structure of multicomponent disordered materials, such as liquid water, aqueous solutions, etc., poses nearly prohibitive difficulties already at the two-particle level, due to the lack of measurable information. For instance, in two-component liquids, like water (composing atoms: O and H) or carbon tetrachloride (composing atoms: C and Cl), there are 3 partial distribution functions (PRDF), g$_{ij}$(r): g$_{\mathrm{OO}}$(r), g$_{\mathrm{OH}}$(r) and g$_{\mathrm{HH}}$(r) for H$_{2}$O and g$_{\mathrm{CC}}$(r), g$_{\mathrm{CCl}}$(r) and g$_{\mathrm{ClCl}}$(r) for CCl$_{4}$. PRDF-s cannot be measured separately (one-by-one) in general; instead, composite functions, called total scattering structure factors \cite{1} (TSSF), F(Q), may be determined in the reciprocal space. The real-space equivalent of F(Q) is its Fourier-transform pair, called the total radial distribution function (TRDF), G(r). F(Q) and G(r) contain contributions from PRDF-s which depend on the scattering power and concentration of the components \cite{2}:
\begin{equation}
\label{GR-def}
G(r) = \sum_{i,j=1} ^{n} c_{i} c_{j} b_{i} b_{j} \left[ g_{i,j}(r)-1\right],
\end{equation}
\begin{equation}
\label{FQ-def}
F(Q) = \rho_{0}\int_{0}^{\infty}4 \pi r^2 G(r)\left( \frac{\sin{Qr}}{Q}\right) \rd r.
\end{equation}

In the above equations, $c_{i}$ and $b_{i}$ are the molar ratio and the neutron scattering length of species $i$, $G(r)$ is the total radial distribution function, $\rho_{0}$ is the number density and $Q$ is the scattering variable (proportional to the {scattering angle); indexes $i$ and $j$ run through nuclear species. We would like to stress that PRDF-s cannot be considered as primary experimental information; it has been known for some time that more than one set of PRDF-s may be equally consistent with a given set of measured structure factors \cite{3}, containing different isotopic ratios of $^{1}$H and D ($^{2}$H) in water. Therefore, any particular set of the so-called  ``experimental'' PRDF-s should be taken only as one possible interpretation of the measured data. That is why the underlying problem of consistency between TSSF-s and PRDF-s deserves particular attention.

Strictly speaking for an unequivocal determination of all the three PRDF-s of any two-component system, three independent measurements would be necessary on the same (at least chemically, if not isotopically, identical) material. This requirement makes the task difficult and performable only for a relatively small number of systems. For this reason, it is of utmost importance to be aware of the quality/reliability of the PRDF-s derived.

Neutron diffraction with isotopic substitution \cite{4,5} (NDIS) and anomalous X-ray scattering \cite{5} allow, in fortunate cases, to separate all the PRDF-s. These fortunate cases are all two-component systems. Arguably the best known example of isotopic substitution is replacing $^{1}$H by $^{2}$H (deuterium). The trick was tried first on water \cite{6}, which liquid has been since probed many times by NDIS (see, e.g.,  \cite{7,8}). The fact that one single experiment has not been able to clarify water PRDF-s for once and forever indicates that there must be some unresolved issues concerning the procedure. Indeed, a most (in)famous property of $^{1}$H nuclei (protons) is their incredibly large incoherent inelastic scattering cross section for neutrons \cite{9}. This property renders more than 95\% of the signal measured by neutron diffraction useless (``background-to-be-subtracted'') from the structural point of view (for a more detailed exposition of the problem, see  \cite{3}).

One of the latest compilation or PRDF-s resulting from H/D NDIS experiments on water can be found in  \cite{10}; it is rather helpful that numerical data for PRDF-s resulting from this work are posted also on the Internet \cite{11}. Due to difficulties mentioned above it would be rather comforting if an independent (and preferably, a positive) assessment of these results could be provided.

In this work, we propose a possible approach. As primary data, recent neutron diffraction results from the same author \cite{8} on liquid heavy water, as well as the most recent published set of X-ray data of Fu et al. \cite{12} will be applied. The procedure uses the so-called Reverse Monte Carlo (RMC) algorithm \cite{13}, based on which one of the present authors (LP) has published, together with Prof. Orest Pizio, a related approach \cite{14}, that later could be applied for assessing the performance of interaction potentials of liquid water \cite{15}.

\section{Reverse Monte Carlo modelling}
Reverse Monte Carlo \cite{13,16,17,18} is a simple tool for constructing large, three-dimensional structural models that are consistent with total scattering structure factors (within the estimated level of their errors) obtained from diffraction experiments. Via random movements of particles, the difference  between experimental and model total structure factors (calculated similarly to the $\chi^{2}$-statistics) is minimized. As a result, by the end of the calculation a particle configuration is available that is consistent with the experimental structure factor(s). If the structure is analyzed further, partial radial distribution functions, as well as other structural characteristics (neighbour distributions, cosine distribution of bond angles) can be calculated from the particle configurations.

A possible algorithm that can realize the above features may be outlined as follows \cite{13}:
\begin{enumerate}
  \item Start with an initial collection of particle coordinates in a cubic box; this may be a crystalline or a random distribution of--at least a few thousands of--particles, or even the final particle distribution from a previous simulation.
  \item Calculate the partial radial distribution functions for the configuration. Compose total radial distribution functions, according to the experimental weighting factors. Use Fourier transformation for calculating total scattering structure factors.
  \item Calculate differences between model and experimental functions as follows (shown here for one single TSSF):
\begin{equation}
\label{CHI2-def}
\chi^{2} [F(Q)] = \sum_{i}{\frac{ \left(F^{\mathrm{C}} \left(Q_{i}\right) - F^{\mathrm{E}} \left(Q_{i}\right)\right)^{2}} { \sigma^{2}}}\,.
\end{equation}
The  ``C'' and  ``E'' superscripts refer to  ``calculated'' and  ``experimental'' functions, respectively; $\sigma$ is a control parameter that is related to the assumed level of experimental errors.
  \item Move one particle at random.
  \item Calculate PRDF-s, TRDF-s, TSSF-s and from them, also the $\chi^{2}$, for the new position.
  \item If the $\chi^{2}$ for the new position is smaller than it was for the old position (i.e., the difference between simulated and measured TSSF-s has become smaller) then accept the move immediately. Otherwise accept the move only with a probability that is proportional to $\exp(-\Delta \chi^{2})$; accepting  ``bad'' moves with such small but finite probability will prevent calculations from sticking in local minima. If a move is  ``accepted'' then the  ``new'' position becomes the  ``old'' one for the next attempted move.
  \item Continue from step 4.
\end{enumerate}

The most valuable feature of the RMC method concerning the present purposes is that it can incorporate any piece of information that can be calculated directly from particle coordinates. Partial radial distribution functions from traditional evaluation of diffraction data  \cite{4,5} fall into this category. In this case, if consistency with all input data is reached then it may be stated that these input data are mutually consistent. If, however, some of the input data cannot be approached within their uncertainties then it means that parts of the input data set are not consistent with other pieces of input information. In our case this would mean that some of the input PRDF-s from the literature  \cite{10} would not be consistent with the experimental input total scattering structure factor(s).

In the RMC calculations that are the basis of the present work, one total scattering structure factor from neutron diffraction  \cite{7,8}, one TSSF from X-ray diffraction \cite{12} and three partial radial distribution functions from  \cite{10,11} are applied as input data for each calculation. The PRDF-s appear for various thermodynamic conditions in  \cite{10,11}; here the ones denoted as set  ``(a)'' are being used. Data for set  ``(a)'' were obtained at 298 K and 0.1 MPa (ambient conditions); the molecular number density was 0.0334 molecules \AA$^{-3}$.

We report two calculations: as real experimental results, the first one uses neutron diffraction data only whereas in the other one, neutron and X-ray TSSF-s are both applied. Naturally, both calculations involve the same set of PRDF-s. In both RMC computations, the simulation box contained 2000 water molecules (6000 atoms). Goodness-of-fit values, $R_{\mathrm{w}}$-s (which are basically sums of the squared differences, see below), are reported in a normalized form, so that variations in terms of the number of $r$ and $Q$ points considered would not affect the assessment; additionally, the applied $r$ and $Q$ ranges were kept as uniform as possible.

For making the definitions of the different $R_{\mathrm{w}}$-s used throughout this study clearer, below we provide the appropriate equations as to how to calculate these differences (for PRDF-s, only the example of the O--O g(r) is shown):
\begin{eqnarray}
\label{RWFQ-def}
{R_{\mathrm{w}}}^2 [F(Q)] &=& \frac{\sum_{i} \left(F^{\mathrm{C}} \left(Q_{i}\right) - F^{\mathrm{E}} \left(Q_{i}\right)\right)^{2} }{ \sum_{i} F^{\mathrm{E}} (Q_{i})^{2}}\,,
\\
\label{RWgr-def}
{R_{\mathrm{w}}}^2 \left[g_{\mathrm{OO}}(r)\right] &=& \frac{\sum_{i} \left(g_{\mathrm{OO}}^{\mathrm{C}} \left(r_{j}\right) - g_{\mathrm{OO}}^{\mathrm{E}} \left(r_{j}\right)\right)^{2}}{ \sum_{i} g_{\mathrm{OO}}^{\mathrm{E}} (r_{j})^{2}}\,.
\end{eqnarray}

In the above expressions, $N_{i}$ and $N_{j}$ are the number of $Q$ and $r$ points, respectively, for the experimental TSSF and  ``experimental''  $g(r)$-s, respectively. Indices ``$C$'' and ``$E$'' refer to ``RMC calculated'' and  ``experimental'' quantities.

\section{Results and discussion}

Table~\ref{table1} summarizes goodness-of-fit values for each calculation reported here. Just by looking at the numbers, it becomes evident that the scheme proposed here looks rather promising by being decisive in a couple of important issues. It is clear that consistently, it is the O--H PRDF that is the least consistent with primary experimental data (the TSSF-s); this is in line with previous findings \cite{3,15}.

\begin{table}[h]
\caption{Goodness-of-fit (R$_{\mathrm{w}}$) values for individual data sets [F$^N$(Q), F$^X$(Q) and the three partial g(r)'s].  ``R$_{\mathrm{w}}$ sum'' is simply the sum of individual R$_{\mathrm{w}}$ values.  (ND: neutron diffraction; XRD: X-ray diffraction.)}
\vspace{1ex}
\label{table1}
\begin{center}
\renewcommand{\arraystretch}{0}
\begin{tabular}{|c|c|c|c|c|}
\hline
  & ND only & ND+XRD \strut\\
\hline\hline
    R$_{\mathrm{w}}$ [ND F(Q)] & 4.67\% &  4.67\% \strut\\
\hline
    R$_{\mathrm{w}}$ [XRD F(Q)] & -- & 4.27\%  \strut\\
\hline
    R$_{\mathrm{w}}$ [g$_{\mathrm{OO}}$(r)] & 4.03\% & 8.79\% \strut\\
\hline
    R$_{\mathrm{w}}$ [g$_{\mathrm{OH}}$(r)] & 16.91\% & 16.93\% \strut\\
\hline
    R$_{\mathrm{w}}$ [g$_{\mathrm{HH}}$(r)] & 3.87\%  & 3.90\% \strut\\
\hline
    R$_{\mathrm{w}}$ sum & -- & 38.56\% \strut\\
\hline
    R$_{\mathrm{w}}$ sum (without XRD) & 29.41\%  & 34.29\%  \strut\\
    \hline
\end{tabular}
\renewcommand{\arraystretch}{1}
\end{center}
\end{table}

It is instructive to look at figures~\ref{fig1} and \ref{fig2}, so that the level of consistency between experimental TSSF-s and PRDF-s may be visualized better. We note here that no separate figures are provided for calculations where only neutron diffraction data were considered, since the tiny differences in terms of the goodness-of-fits are not visible, apart from the O--O PRDF, which is discussed below more in detail.
\begin{figure}[ht]
\includegraphics[width=0.48\textwidth]{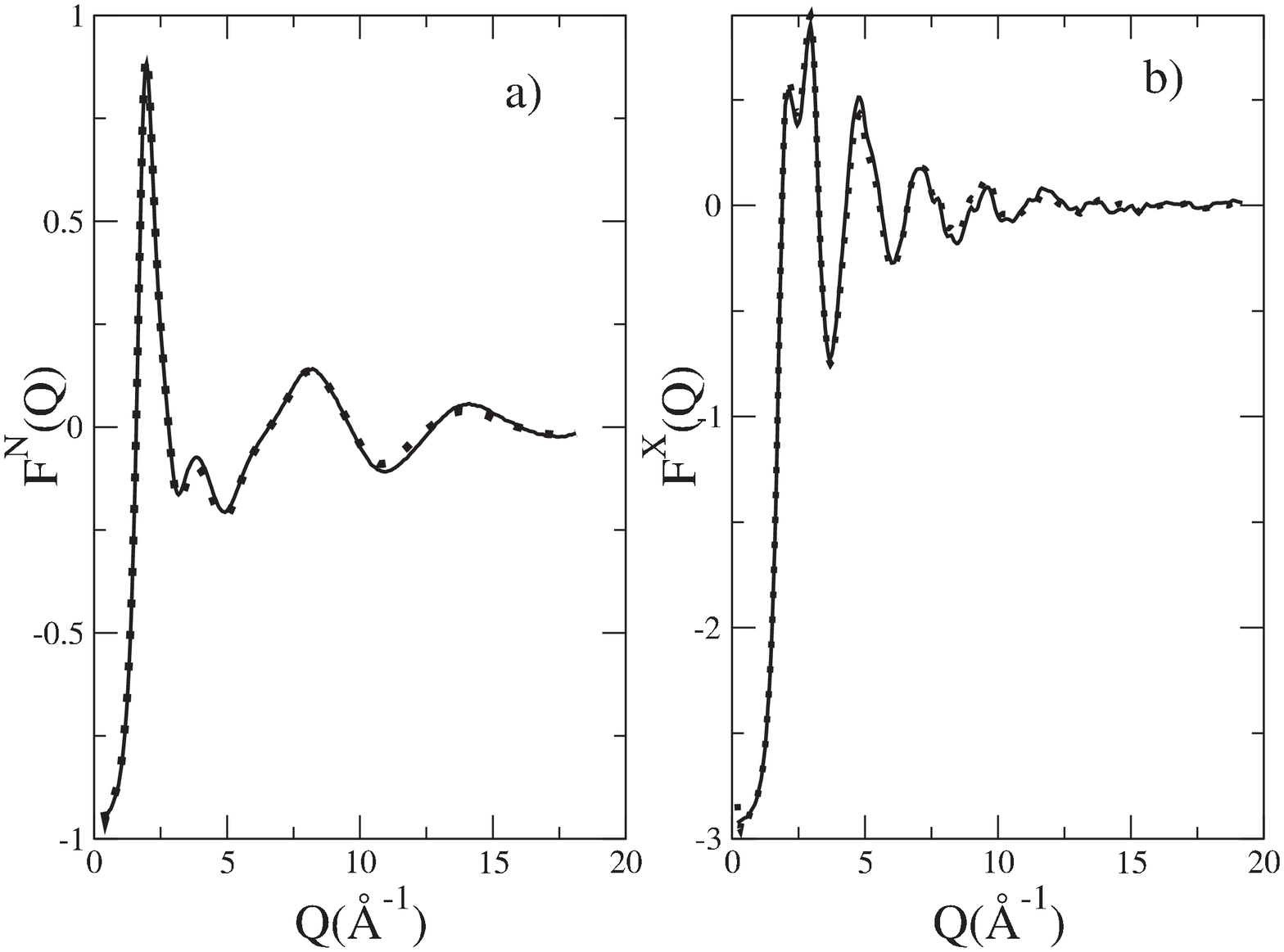}%
\hfill%
\includegraphics[width=0.48\textwidth]{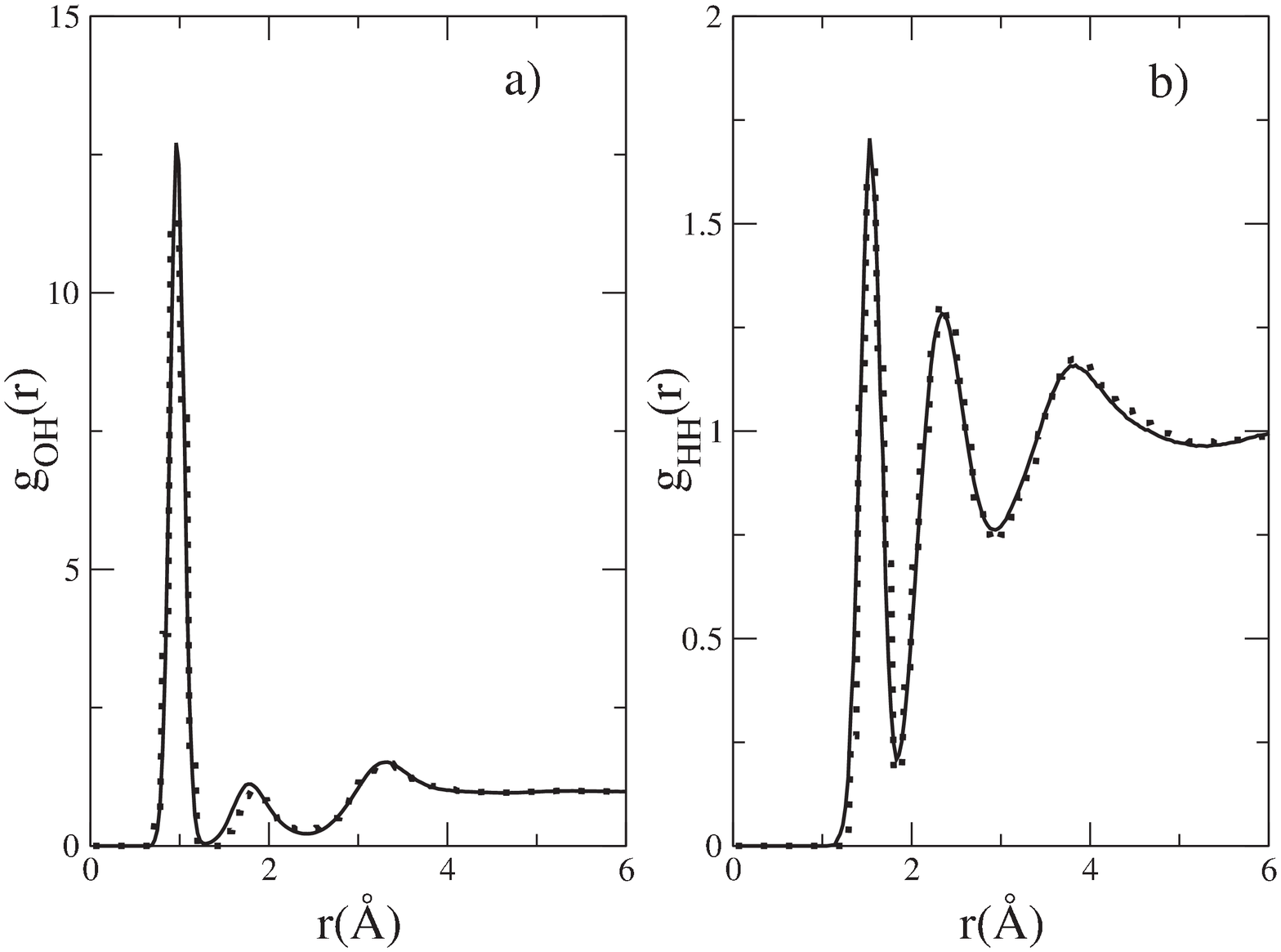}%
\\%
\parbox[t]{0.48\textwidth}{%
\caption{%
RMC modelling  ``experimental'' partial radial distribution functions of liquid water (set  ``(a)'' in  \cite{10,11}), together with neutron diffraction TSSF of heavy water \cite{7,8} (part {\textit a}) and X-ray diffraction TSSF of light water \cite{12} (part {\textit b}). Solid line: experiment; dotted line: RMC.}
\label{fig1}%
}%
\hfill%
\parbox[t]{0.48\textwidth}{%
\caption{%
RMC modelling  ``experimental'' O--H (part {\textit a}) and H--H (part {\textit b}) partial radial distribution functions of liquid water (set (a) in  \cite{10,11}), together with neutron diffraction TSSF of heavy water \cite{7,8} and X-ray diffraction TSSF of light water \cite{12}. Solid line:  ``experiment'' (  \cite{11}); dotted line: RMC.}
\label{fig2}%
}%
\end{figure}

Even though in each calculation the R$_{\mathrm{w}}$ factor for the O--H PRDF is significantly higher than that for the other two partials, agreement between RMC and  ``experiment'', and consequently, between primary experimental information and PRDF-s derived from it in  \cite{10}, is rather convincing.
\begin{figure}[!h]
\centerline{\includegraphics[width=8cm]{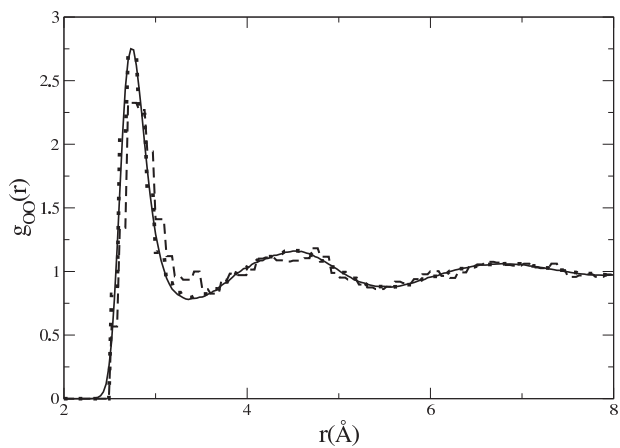}}
\caption{RMC modelling the  ``experimental'' O--O partial radial distribution function of liquid water. Solid line: g$_{\mathrm{OO}}$(r) from set (a) in  \cite{10,11}); dotted line: RMC model when applying neutron diffraction data on heavy water \cite{7,8} ONLY; dashed line: RMC model when applying both neutron and X-ray diffraction TSSF-s.}
\label{fig3}
\end{figure}

\looseness=-1 It is obvious from both table~\ref{table1} and figure~\ref{fig3} that the introduction of X-ray diffraction data influences the O--O PRDF the most. Clearly, the quantitative consistency with the neutron data (i.e., where the calculation did NOT contain X-ray data), has worsened substantially. This finding indicates that PRDF-s derived from neutron diffraction data with isotopic substitution are not entirely consistent with the latest (and independent) set of X-ray TSSF. Note also that the X-ray data are dominated by the O--O PRDF (whereas to the neutron weighted TSSF of D$_{2}$O, the contribution is less than 10 \%, see   \cite{3,7,8,10}), so that--provided that errors in the two TSSF-s are comparable--in terms of O--O correlations, the X-ray TSSF should be considered as decisive. This is a straightforward indication that further improvements would be needed to the procedures applied during data evaluation of neutron diffraction data sets that work with H/D isotopic substitution.

Given the simplicity of the approach, its application might be fruitful in other areas, too. For example, statistical theories have also been frequently used for producing PRDF-s of various liquids (see, e.g.  \cite{19,20} and references therein). To assess the consistency level of these theories with experimental data, the RMC-based calculations described above might be adequate.

\section{Conclusions}

A simple approach has been devised in order to establish the level of consistency between experimental diffraction data, that can be obtained in the reciprocal space in the form of composite total scattering structure factors, and partial radial distribution functions that may be derived from them.

For the particular case of the most frequently cited liquid water PRDF-s  \cite{10} it has been established that they are consistent with the most reliable neutron diffraction data set, taken on pure heavy water \cite{7,8}. On the other hand, the O--O partial radial distribution function provided by  \cite{10} is not entirely in agreement with the results of the latest X-ray diffraction measurements  \cite{12} on liquid water.

\section*{Acknowledgements}

This work has been partially supported by the Hungarian Basic Research Fund (OTKA), Grant No.~K083529.

\newpage

\ukrainianpart

\title{Незалежний, загальний метод для перевірки узгодження дифракційних
даних з парціальними радіальними функціями розподілу, які отримані з
них: приклад рідкої води}

\author{Ж.~Штайнцінґер\refaddr{label1}, Л.~Пустаі\refaddr{label2}}

\addresses{
\addr{label1} Школа другого ступеня Будаі Надь Антал, H-1121,
Будапешт, Угорщина
\addr{label2} Інститут фізики і оптики твердого
тіла,
Віґнерівський дослідний центр фізики, Угорська Академія наук, H-1525 Будапешт,
Угорщина}

\makeukrtitle

\begin{abstract}
\tolerance=3000%
Існує декілька шляхів отримання парціальних радіальних функцій
розподілу в невпорядкованих системах із експериментальних
дифракційних (і/або EXAFS) даних. Через обмеженість та похибки
експериментальних даних, як і недосконалість обчислювальних
процедур, першочергової важливості набуває підтвердження того, що
кінцеві результати (парціальні радіальні функції розподілу) і первинна
інформація (дифракційні дані) узгоджуються між собою. Пропонується
простий підхід, який базується на оберненому моделюванні Монте
Карло, який спроможний розв'язати цю дилему. В якості демонстрації
ми використовуємо найцитованіший набір ``експериментальних''
парціальних радіальних функцій розподілу для води і досліджуємо, чи усі
три парціальні розподіли (O--O, O--H і H--H) узгоджуються з повним
структурним фактором чистої води H$_{2}$O, отриманим із дифракції
X-променів. Ми показуємо, що хоча дані нейтронного розсіювання на
важкій воді цілком відповідають усім парціальним розподілам,
додаткове врахування даних розсіяння X-променів виявляє
проблеми з парціальною функцією розподілу O--O. Ми пропонуємо
застосовувати запропонований тут підхід також для вияснення того, чи
парціальні радіальні функції розподілу, які отримані із статистичних
теорій рідкого стану, узгоджуються із виміряними структурними
факторами.
\keywords нейтронне розсіяння, парціальні радіальні функції розподілу,
обернене моделювання Монте Карло

\end{abstract}

\end{document}